\documentclass[pra,aps,twocolumn,nopacs,superscriptaddress,nofootinbib]{revtex4}
\usepackage{graphicx}
\usepackage{dcolumn}
\usepackage{bm}
\usepackage{amsmath}
\usepackage{epsfig}
\usepackage{braket}
\usepackage{slashbox}
\usepackage{bbm}
\usepackage{color}
\usepackage{braket}

\begin{document}

\def\ii{{\rm i}}  \def\ee{{\rm e}}  \def\vb{{\bf v}}
\def\Rb{{\bf R}}  \def\pb{{\bf p}}  \def\Qb{{\bf Q}}  \def\ub{{\bf u}}  \def\db{{\bf d}}
\def\jb{{\bf j}}  \def\kb{{\bf k}}  \def\Eb{{\bf E}}  \def\rb{{\bf r}}
\def\xx{\hat{\bf x}}  \def\yy{\hat{\bf y}}  \def\zz{\hat{\bf z}}
\def\th{\vec{\bf{\theta}}}  \def\wp{{\omega_{\rm p}}}
\def\fE{\vec{\mathcal{E}}}  \def\Mb{{\bf M}}  \def\EF{{E_{\rm F}}}
\def\vF{{v_{\rm F}}}   \def\kF{{k_{\rm F}}}   \def\EFT{{E_{\rm F}^T}}
\def\Jb{{\bf J}} \def\eb{{\bf e}}  \def\HTB{H_{\text{TB}}}

\title{Plasmon-assisted high-harmonic generation in graphene}
\author{Joel~D.~Cox}
\email{joel.cox@icfo.es}
\affiliation{ICFO-Institut de Ciencies Fotoniques, The Barcelona Institute of Science and Technology, 08860 Castelldefels (Barcelona), Spain}
\author{Andrea~Marini}
\email{andrea.marini@icfo.es}
\affiliation{ICFO-Institut de Ciencies Fotoniques, The Barcelona Institute of Science and Technology, 08860 Castelldefels (Barcelona), Spain}
\author{F.~Javier~Garc\'{\i}a~de~Abajo}
\email{javier.garciadeabajo@icfo.es}
\affiliation{ICFO-Institut de Ciencies Fotoniques, The Barcelona Institute of Science and Technology, 08860 Castelldefels (Barcelona), Spain}
\affiliation{ICREA-Instituci\'o Catalana de Recerca i Estudis Avan\c{c}ats, Passeig Llu\'{\i}s Companys 23, 08010 Barcelona, Spain}

\begin{abstract}
{\bf
High-harmonic generation (HHG) in condensed-matter systems is both a source of fundamental insight into quantum electron motion and a promising candidate to realize compact ultraviolet and ultrafast light sources. Here we argue that the large light intensity required for this phenomenon to occur can be reached by exploiting localized plasmons in conducting nanostructures. In particular, we demonstrate that doped graphene nanostructures combine a strong plasmonic near-field enhancement and a pronounced intrinsic nonlinearity that result in efficient broadband HHG within a single material platform. We extract this conclusion from time-domain simulations using two complementary nonperturbative approaches based on atomistic one-electron density-matrix and massless Dirac-fermion Bloch-equation pictures. High harmonics are predicted to be emitted with unprecedentedly large intensity by tuning the incident light to the localized plasmons of ribbons and finite islands. In contrast to atomic systems, we observe no cutoff in harmonic order. Our results support the strong potential of nanostructured graphene as a robust, electrically tunable platform for HHG.}
\end{abstract}
\maketitle
% {\bf KEYWORDS:} graphene, plasmons, nonlinear response, optical tunability

\section*{Introduction}

High-harmonic generation (HHG) is an extreme nonlinear optical phenomenon first observed by driving atomic gases with intense ultrashort light pulses \cite{FLL1988, LB93_2}. The harmonic intensity remains surprisingly large up to a high order of the pulse carrier frequency, stimulating applications for HHG as a source of ultraviolet and x-ray radiation \cite{SBS97_2,BK00,GSH08}, as well as in the generation of attosecond pulses \cite{PTB01,CK07,KI09}, which has enabled tomographic imaging of molecular orbitals \cite{ILZ04} and the exploration of subfemtosecond dynamics in chemical reactions \cite{WBK10}.

Recent observations of HHG from condensed-matter systems \cite{GDS11,SHL14,LGK15,VHT15,HLS15} are currently attracting much interest not only in the pursuit of new solid-state optical technologies, but also in the underlying physics of HHG in bulk crystals and its analogy with atomic gases. Indeed, while HHG from individual atoms is well-understood as the coherent emission produced by the optically induced tunneling ionization of an electron, its acceleration by the driving field, and the subsequent recollision with its parent ion \cite{C93_2,LBI94}, the picture becomes less clear in crystalline media, where collective effects associated with the high density of electrons and their interaction with the lattice significantly complicate the generation process. As expected, HHG in solids is found to depend strongly on the electronic band structure and the interplay between inter- and intraband transitions \cite{GDS11,SHL14,VHT15,TIO16,OCO16}.

The linear, gapless dispersion relation of graphene electrons \cite{W1947,GN07} garners strong interest in the nonlinear optical response of the atomically thin material, which recent experiments demonstrate to be intrinsically large \cite{HHM10,WZY11,GPM12,KKG13,HDP13,ARD14}. On the theory side, monolayer graphene is expected to produce intense HHG in the THz regime \cite{I10,ASD14}, attributed to complementary inter- and intraband charge carrier motion at low temperatures and doping levels. Unfortunately, recent experiments report either no evidence \cite{PCT13} or only a weak effect \cite{BMR14} associated with the generation of low-order harmonics from multilayer graphene for currently available THz illumination intensities. This situation could be improved by using more intense sources at higher frequencies, and further relying on enhanced graphene-light interaction mediated by localized plasmon resonances.

Graphene plasmons \cite{JBS09,FAB11,paper176,GPN12,paper235}, which provide an efficient way to couple the carbon layer with impinging light, are capable of generating intense local electric fields that are essential to trigger nonlinear optical phenomena. This near-field enhancement, in combination with the highly anharmonic response of graphene \cite{M07_2,ASD14,M16}, is predicted to give rise to large optical nonlinearies \cite{paper247,JC15,CVS15,CYJ15,paper269}. Importantly, these plasmons only exist in highly doped graphene, while their frequency is strongly dependent on the doping level \cite{JBS09,FAB11,paper176,GPN12,paper235}. Electrical gating thus provides a mechanism to tune the harmonic generation in graphene to the desired frequency range.

Here we predict that highly efficient HHG takes place in doped graphene nanostructures when the incident light is tuned to their localized plasmons. Specifically, we obtain harmonic intensities that are orders of magnitude higher than in other materials. Additionally, no sharp cutoff is observed with harmonic order. Our results are based on nonperturbative time-domain numerical simulations of the nonlinear optical response of graphene using two complementary approaches: a random-phase approximation (RPA) description of the single-particle density matrix within a tight-binding (TB) model for the electrons of ribbons and finite islands \cite{paper247}; and the solution of the single-particle Bloch equations for massless Dirac-fermions (MDFs) in extended graphene, complemented by a classical electromagnetic (CEM) description of the self-consistent field produced by the illuminated nanostructure (see Methods). We find both approaches to be in excellent agreement at intensities below the saturable absorption threshold. Our prediction of highly efficient HHG assisted by coupling to graphene plasmons suggests applications to a wide range of nonlinear photonic technologies, including tunable sources of broadband attosecond light.

\section*{Results}

\begin{figure*}[t]
\includegraphics[width=0.9\textwidth]{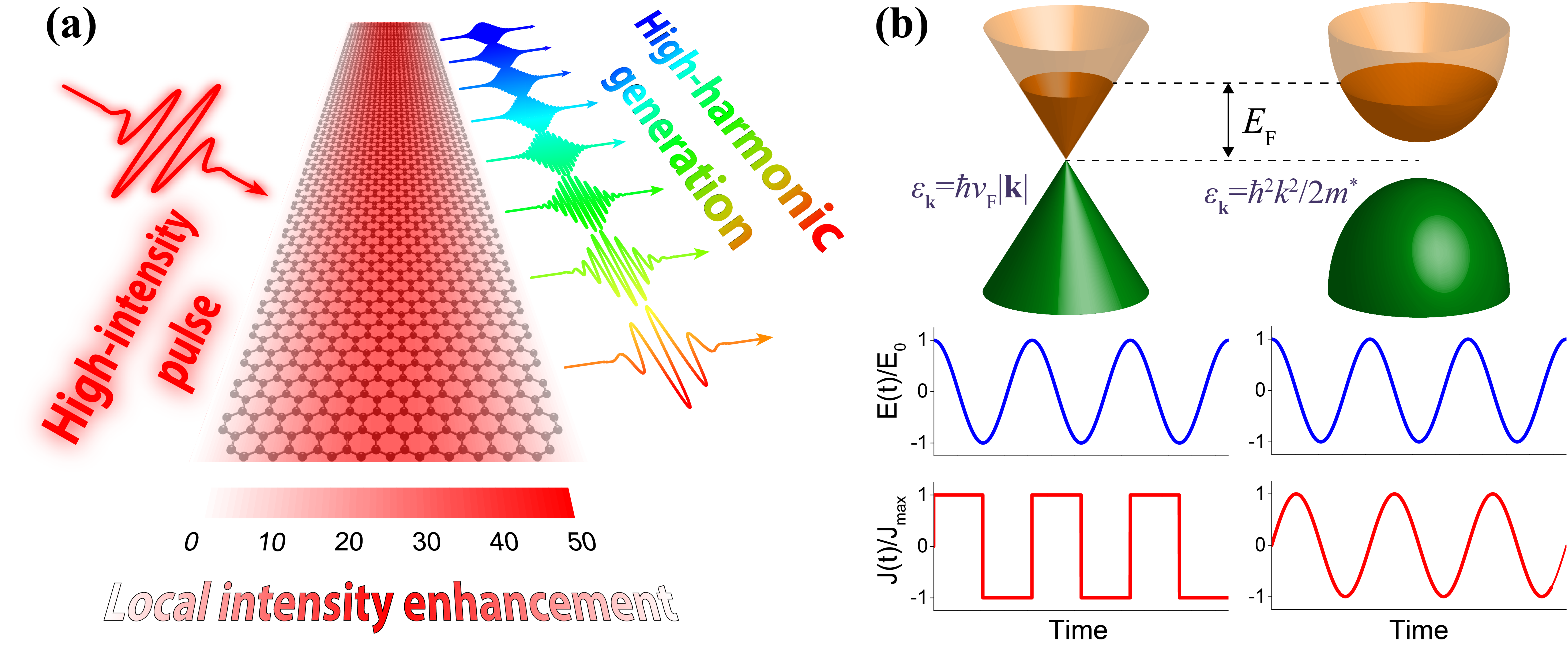}
\caption{{\bf High-harmonic generation (HHG) assisted by graphene plasmons} {\bf (a)} Schematic illustration of a doped graphene nanoribbon illuminated by an intense optical pulse that is resonant with the ribbon transverse dipole plasmon. The latter produces strong in-plane electric-field intensity enhancement (see color scale) that boosts the generation of high harmonics. {\bf (b)} The low-energy band structures of graphene (upper left) and a conventional 2D semiconducting crystal (upper right) respond differently to a monochromatic light electric field $E(t)=E_0\cos(\omega t)$: in graphene, the induced current $J(t)$ (lower left) acquires a square-wave temporal profile, which contains all odd-order harmonics in its Fourier decomposition, while the semiconductor responds harmonically at the driving frequency $\omega$. A 2D free electron gas also shows a harmonic response \cite{BJH1989}.} \label{Fig1}
\end{figure*}

In practice, cumbersome laser amplification schemes are usually needed to reach the extreme electromagnetic field intensities required to generate high-order harmonics. To overcome this limitation, plasmonic nanostructures have attracted considerable interest as {\it{in situ}} electric field enhancers for HHG in gaseous media \cite{KJK08,PKC11,HIH11,TYC16}. As illustrated schematically in Fig.\ \ref{Fig1}a, we propose that compact, efficient HHG can be realized in graphene by combining the intense near-field enhancement associated with graphene plasmons with the intrinsically high nonlinear optical response of this material. The appeal of graphene as a nonlinear optical material stems in part from its linear charge carrier dispersion with electron wave vector $\kb$ at low energies, $\varepsilon_\kb=\hbar\vF|\kb|$, where $\vF\approx c/300$ is the Fermi velocity. In the single-particle MDF description of doped monolayer graphene, neglecting interband electronic transitions, this linear dispersion relation leads to a maximum achievable surface current density $J_{\text{max}}=-en\vF\,{\rm sign}\{\sin(\omega t)\}$ when illuminated by a monochromatic field $E(t)=E_0\cos(\omega t)$ in the $E_0\rightarrow\infty$ limit \cite{M07_2,I10}. The current is thus limited by the doping charge-carrier density $n$. This square-wave profile of the induced current density under intense illumination translates into efficient generation of odd-ordered harmonics (see Fig.\ \ref{Fig1}b). Conversely, in conventional 2D media, for which charge carriers obey a parabolic dispersion relation $\varepsilon_\kb=\hbar^2 k^2/2m^*$, the system responds harmonically at the frequency of the driving field, regardless of electron-electron interactions \cite{BJH1989}. While this comparison favorably portrays graphene as a highly nonlinear optical material, it is important to note that interband optical transitions compensating the large intraband anharmonicity become significant at high intensities, even when the system is driven at frequencies below the Fermi level \cite{I10}.

\begin{figure*}[t]
\includegraphics[width=1\textwidth]{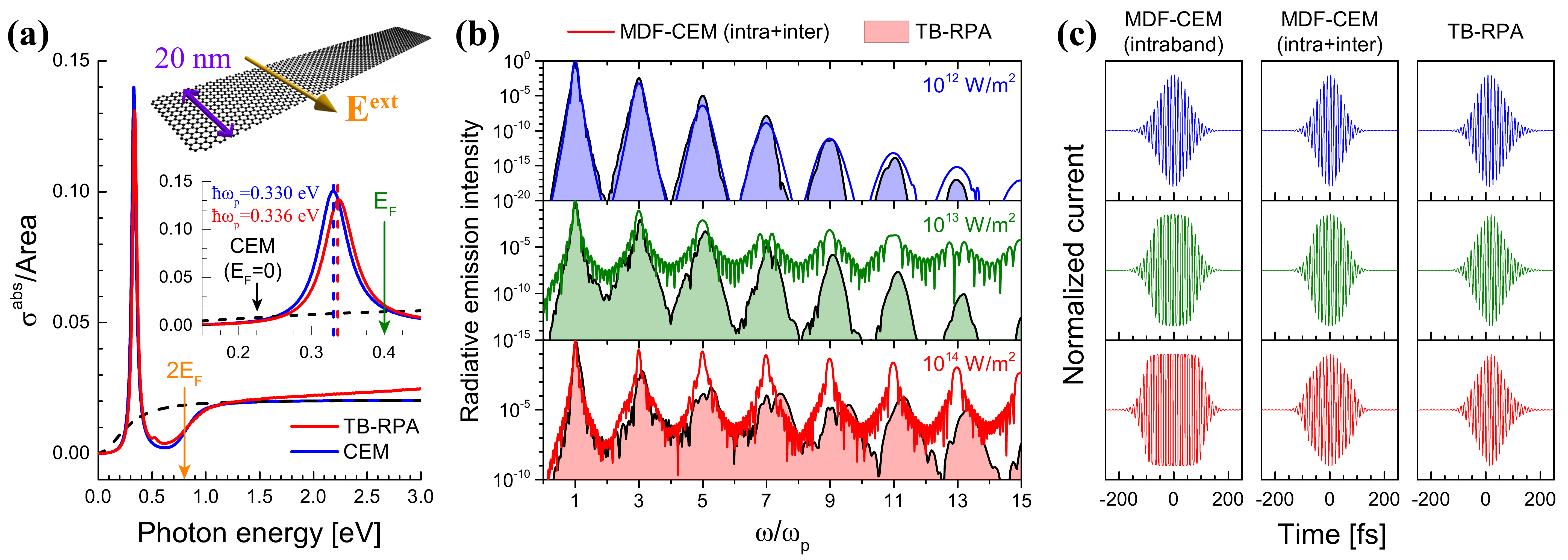}
\caption{{\bf HHG from a graphene nanoribbon.} {\bf (a)} Absorption cross-section of a 20-nm-wide armchaired graphene nanoribbon doped to a Fermi energy $\EF=0.4$\,eV, as predicted by atomistic (TB-RPA, red curves) and classical electrodynamic (local-RPA conductivity at $T=300\,$K, blue curves) simulations for transverse light polarization (see upper inset). The prominent resonant feature (see detail in the central inset) corresponds to the transverse dipolar plasmon within the $2\EF$ optical gap. The dashed curve shows the classical simulation for the undoped ribbon. {\bf (b)} Spectral decomposition of the light emission energy under illumination by a normally incident pulse (100\,fs FWHM duration, centered at the frequency $\wp$ of the ribbon plasmon), as calculated in the time-domain for three different pulse peak intensities (see legend) within the atomistic TB-RPA (filled curves, $\hbar\wp=0.336$\,eV) and MDF-CEM (unfilled curves, $\hbar\wp=0.330$\,eV) descriptions. Each curve is normalized to its own maximum value around the fundamental frequency. {\bf (c)} Temporal evolution of the induced currents corresponding to the plots in (b).}
\label{Fig2}
\end{figure*}

Quantitative analysis of plasmon-enhanced HHG in a doped graphene nanoribbon is presented in Fig.\ \ref{Fig2}. The linear optical absorption of the nanoribbon under consideration (20\,nm width, $\EF=0.4$\,eV Fermi energy) shows a prominent dipolar plasmon (Fig.\ \ref{Fig2}a), as predicted by TB-RPA atomistic simulations and classical electrodynamics, in excellent mutual agreement. We thus consider HHG produced by incident pulses with central frequency tuned to that plasmon. We present HHG simulations obtained with the MDF-CEM and TB-RPA approaches (see Methods) in Fig.\ \ref{Fig2}b, which shows the spectral decomposition (time-Fourier transform) of the radiative emission intensities for 100\,fs incident light pulses with three different peak intensities. Each spectrum is normalized to the maximum value around the fundamental frequency. The corresponding temporal evolution of the graphene induced current is shown in Fig.\ \ref{Fig2}c. Remarkably, high harmonics up to $13^\text{th}$ order are clearly discernible in the emission spectrum even at a relatively low incident peak intensity $I_0=10^{12}$\,W/m$^2$. The agreement between MDF-CEM and TB-RPA descriptions is then excellent both in the spectra (Fig.\ \ref{Fig2}b, upper plots) and in the time-resolved induced current (Fig.\ \ref{Fig2}c). The temporal evolution of the induced current tends to follow the profile of the incident Gaussian pulse, although a small time delay of the peak current is observed in the atomistic simulation due to the self-consistent Coulomb interaction, which persists beyond the duration of the pulse on a timescale determined by the inelastic relaxation time $\tau=13.2\,$fs. By raising the peak intensity, the conversion efficiency of high-order harmonics drastically increases in the MDF-CEM picture, while a more modest, yet impressive, enhancement is predicted in the atomistic TB-RPA simulations. Finite-size effects that are included in the atomistic simulations but not in the MDF-CEM description (see Methods) contribute to this discrepancy. Additionally, the plasmonic local-field enhancement is  self-consistently described in the TB-RPA approach, but not in the MDF-CEM method. For the high level of doping under consideration, intraband electronic transitions dominate the optical response, particularly at low intensities, while interband transitions reduce the level of anharmonicity, as observed in the temporal profiles of the induced current when comparing MDF-CEM simulations with (center plots) and without (left plots) inclusion of interband processes (Fig.\ \ref{Fig2}c). 

\begin{figure*}[t]
\includegraphics[width=1.0\textwidth]{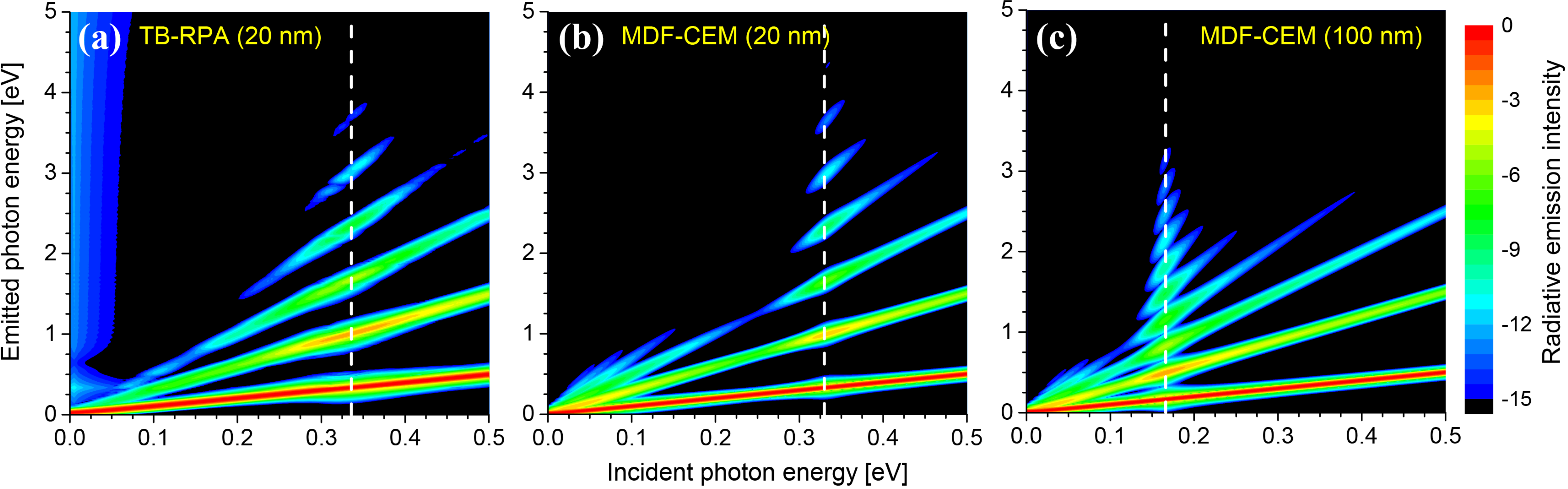}
\caption{{\bf Dependence of HHG on incident photon energy in graphene nanoribbons.} We show the emission intensity from doped graphene nanoribbons under transverse normal illumination as a function of the incident and emitted photon energies as calculated within the (a) TB-RPA and (b,c) MDF-CEM approaches. The incident pulse has a FWHM duration of 100\,fs and a peak intensity of $10^{12}$\,W/m$^2$, while the Fermi energy is 0.4\,eV in all cases. The ribbon width is 20\,nm in (a,b) and 100\,nm in (c), giving rise to the plasmon energies indicated by the vertical dashed lines.} \label{Fig3}
\end{figure*}

The dramatic increase in HHG from localized plasmons in graphene nanoribbons is clearly shown in Fig.\ \ref{Fig3} by mapping the emission intensity over a wide range of input pulse carrier frequencies, where at each input frequency the response is normalized to its respective maximum at the fundamental harmonic. Noticeable enhancement in harmonic generation appears when the excitation frequencies coincide with the plasmon resonance, which can be tuned actively via electrostatic gating and passively by selecting different ribbon widths. Although yet high-order harmonics appear in the spectra, we restrict our investigation to low photon energies where the tight-binding model for graphene remains valid (i.e., below the $\pi$ plasmon near $5$\,eV). In Fig.\ \ref{Fig3}a,b we present results for the doped 20\,nm ribbon considered previously, based on atomistic TB-RPA and MDF-CEM simulations, respectively, for 100\,fs pulses with $10^{12}$\,W/m$^2$ peak intensity as those considered in the upper panel of Fig.\ \ref{Fig2}b. While atomistic simulations quickly become computationally unaffordable for ribbons wider than a few tens of nanometers, the MDF-CEM approach enables the exploration of HHG in much larger structures, such as the 100\,nm-wide ribbon explored in Fig.\ \ref{Fig3}c, which is found to generate plasmon-enhanced high-order harmonics with superior efficiency than the 20\,nm ribbons. The red-shifted plasmon resonances found in larger graphene nanostructures naturally lead to higher optical nonlinearities due to their increased proximity to the Dirac point \cite{I10}.

\begin{figure*}[t!]
\includegraphics[width=0.8\textwidth]{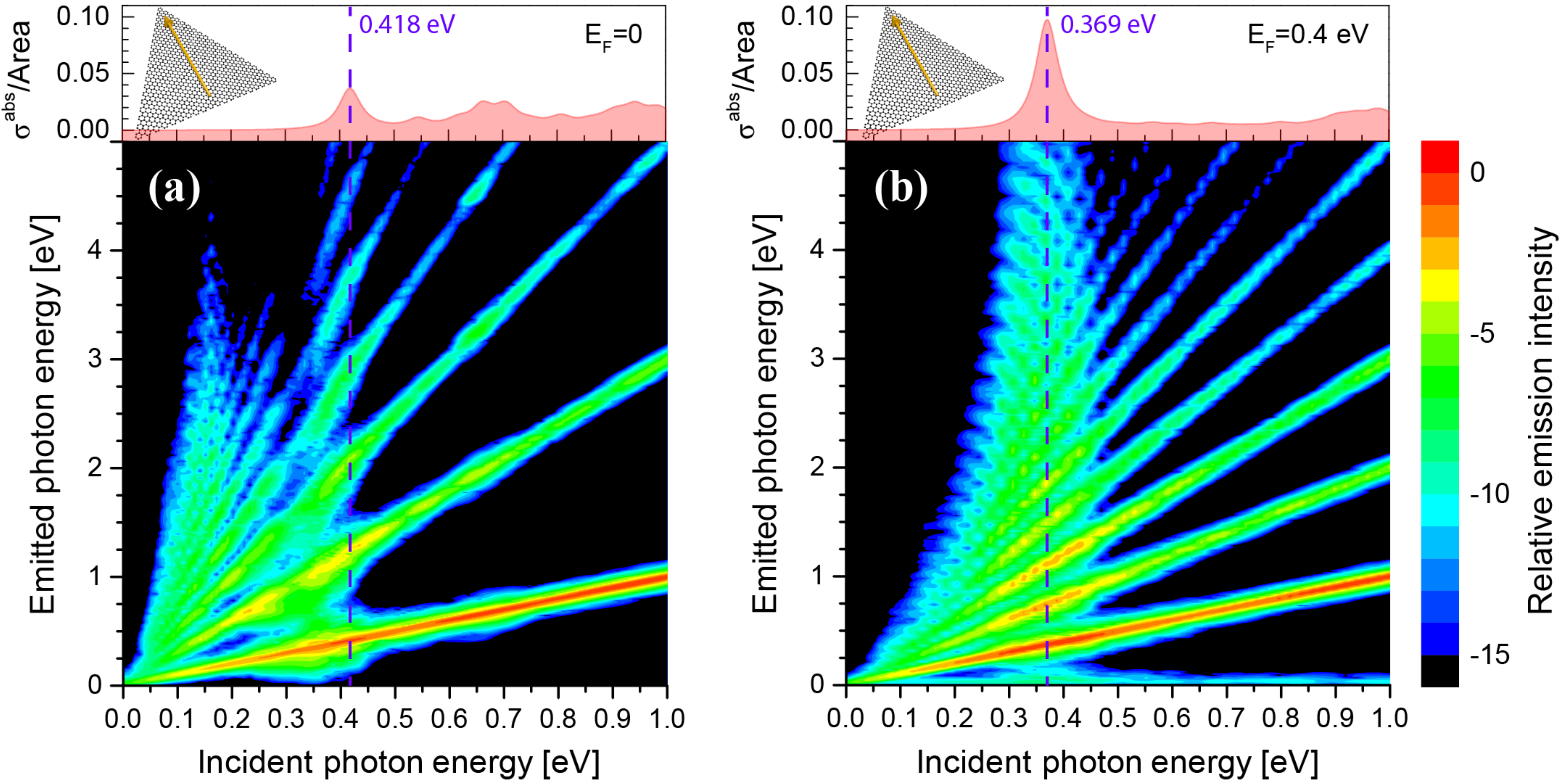}
\caption{{\bf Nonlinear optical response of triangular nanographenes.} We show the emission intensity from doped graphene nanotriangles (armchair edges, equilateral 15\,nm side length) as a function of incident and emitted photon energies, calculated within the TB-RPA approach upon pulse irradiation (100\,fs FWHM duration, $10^{12}$\,W/m$^2$ peak intensity). We consider both (a) undoped and (b) doped ($\EF=0.4$\,eV) triangles. The upper plots show the linear absorption spectrum and the orientation of the normally incident light polarization (insets).} \label{Fig4}
\end{figure*}

Although graphene possesses a centrosymmetric crystal lattice, the geometry of a finite nanostructure can be chosen in a manner that breaks inversion symmetry, enabling even-ordered nonlinear response in certain directions. In Fig.\ \ref{Fig4} we present atomistic TB-RPA simulations of HHG in an armchair-edged 15\,nm equilateral graphene nanotriangle for incident light polarized normal to one of the triangle sides. When the nanotriangle is doped to a Fermi energy $\EF=0.4$\,eV and illuminated with pulses resonant with the dominant, low-energy plasmon mode (Fig.\ \ref{Fig4}b), high harmonics of both even and odd orders are generated with a similar efficiency to the previously considered graphene nanoribbon (cf. Figs.\ \ref{Fig3}a and \ref{Fig4}b). Despite the inversion symmetry of the atomic lattice, a nonzero even-order nonlinear current is produced by a combination of the strong local-field-intensity gradient and the relatively high Fermi wavelength $\lambda_{\rm F}\sim10\,$nm \cite{paper259}, which is commensurate with the size of the triangle. In contrast, only odd-ordered harmonics appear if the nanoisland is undoped (Fig.\ \ref{Fig4}a), as both of these effects (field gradient and long $\lambda_{\rm F}$) are then absent.

\begin{figure*}[t!]
\includegraphics[width=0.9\textwidth]{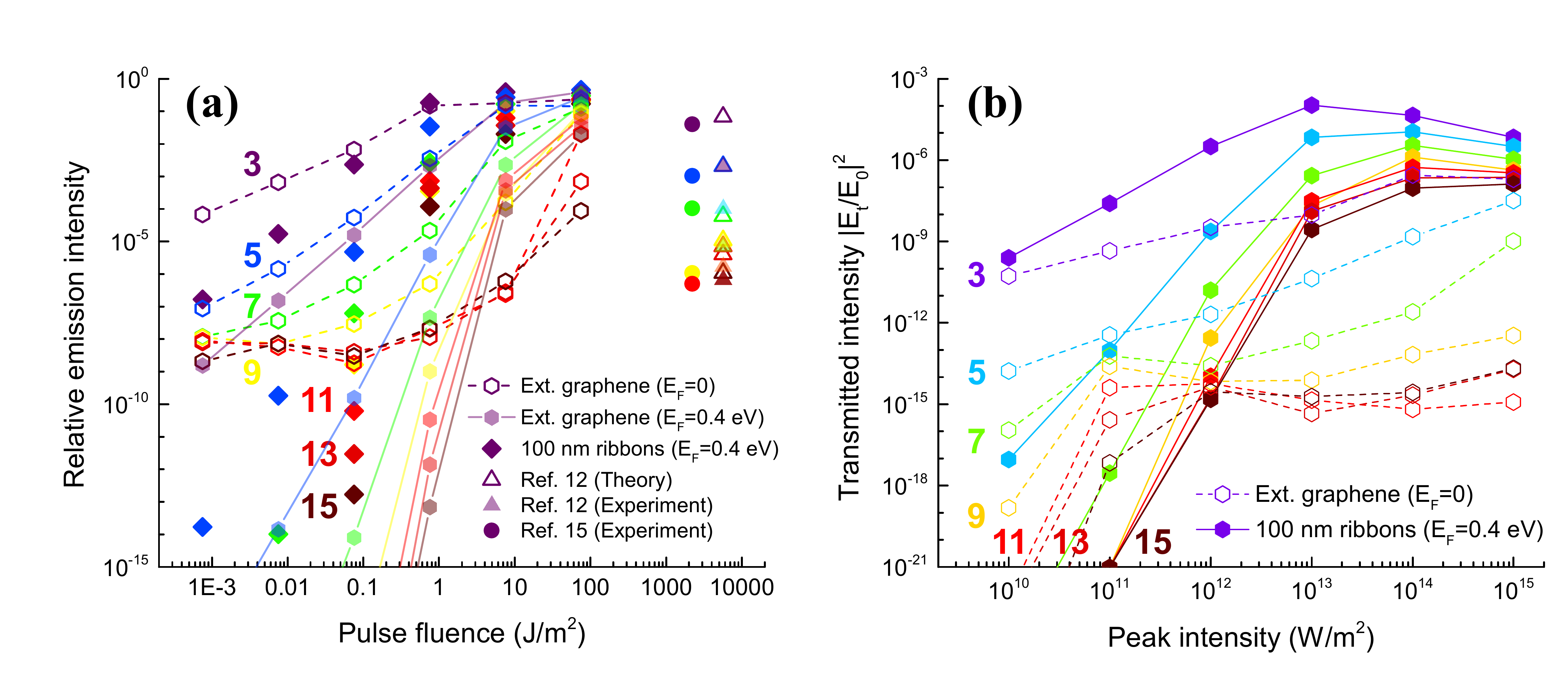}
\caption{{\bf Comparison of graphene plasmon-assisted HHG with extended graphene and measured bulk semiconductors.} {\bf (a)} We show the emission intensities of high harmonics normalized to the intensity of the fundamental peak as a function of pulse fluence. The harmonic index is indicated by the color-coordinated symbols and numbers. We present MDF-CEM simulations for a 100\,nm-wide graphene nanoribbon doped to $E_F=0.4$\,eV (filled diamonds) along with undoped ($E_F=0$, open hexagons) and doped ($E_F=0.4$\,eV, filled hexagons) extended graphene, which we compare with theoretical (open symbols) and experimental (solid symbols) results for bulk GaSe taken from Refs. \cite{SHL14} (triangles) and \cite{HLS15} (circles). The pulses (100\,fs FWHM duration, 0.158\,eV peak energy) are resonant with the ribbon plasmon. {\bf (b)} Comparison of HHG transmission intensity normalized to the incident intensity for a doped-graphene ribbon array (100\,nm width, 200\,nm period, $\EF=0.4\,$eV, solid diamonds) and undoped extended graphene (open hexagons). The intensities of transmitted ($\propto|E_{\rm t}|^2$) to incident ($\propto|E_0|^2$) light are taken at their peak frequencies.} \label{Fig5}
\end{figure*}

Ultimately, we are interested in producing intense high harmonics with moderate incident intensities. With this goal in mind, we analyze the performance of graphene for HHG in Fig.\ \ref{Fig5} and also compare the results with available experiments in solid state systems. As a first observation, even without the involvement of plasmons, the strong intrinsic nonlinearity of graphene is capitalized in a large relative intensity of high harmonics normalized to the response at the fundamental frequency (Fig.\ \ref{Fig5}a): the relative harmonic emission reaches the values measured in GaSe samples, but using 3-4 orders of magnitude lower pulse fluence. It should be noted that a level of theory similar to the MDF model produces excellent agreement with experiment in GaSe (cf. open and solid triangles in Fig.\ \ref{Fig5}a), thus supporting the predictability of our results, which is also emphasized by the agreement between MDF-CEM and atomistic simulations shown in Figs.\ \ref{Fig2} and \ref{Fig3}. By patterning the graphene into ribbons and tuning the incident light to the dominant dipole plasmon energy, HHG is boosted even more, a result that is particularly evident when analyzing the absolute harmonic intensity of resonant ribbons and extended graphene (Fig.\ \ref{Fig5}b). Incidentally, in contrast to the enhancement observed in doped ribbons by exciting the plasmons, doping is detrimental in extended graphene because the Fermi level is then situated in a region where the difference between parabolic and linear electronic band dispersions is reduced, and so is the nonlinear response.

\section*{Conclusion}

In summary, we predict that the combination of high intrinsic nonlinearity and strong plasmonic field confinement provided by doped graphene nanostructures under resonant illumination leads to unprecedentedly high HHG conversion efficiencies. Despite the fact that this material is only one atom thick, we show that it outperforms other solid state systems, such as GaSe, for which HHG measurements have been reported. It should be noted that our results are based on a conservative value of the phenomenological electronic relaxation time $\tau$. The availability of high-quality graphene samples, in which $\tau$ is an order of magnitude longer, should boost HHG in this material even further. We have focused on relatively low fundamental frequencies, so that the high harmonic energies under consideration still lie within a range for which the optical response is dominated by the $\pi$ band of graphene. At low intensities, the response is well described by the low-energy, linear-dipersion region of the electronic band, which explains the agreement that we find between continuum DFM-CEM and atomistic TB-RPA descriptions. Although future work is required to extend these results to higher photon energies, which will require the involvement of deeper electron bands, we conclude the HHG conversion efficiencies associated with localized plasmons in graphene nanostructures appear to be remarkably high for an atomic layer, indicating a strong potential for developing electrically tunable, ultra-compact nonlinear photonic technologies.

\section*{Methods}

\subsection{TB-RPA simulations}

We follow a previously-reported atomistic approach \cite{paper183,paper247,paper269} to simulate the nonlinear optical response of graphene nanostructures via direct time-domain integration of the single-electron density matrix equation of motion,
\begin{equation}
\dot{\rho} = -\frac{\ii}{\hbar}[\HTB-e\phi,\rho]-\frac{1}{2\tau}\left(\rho-\rho^0\right),
\nonumber
\end{equation}
where $\HTB$ is a tight-binding Hamiltonian describing the one-electron states of the $\pi$ band of graphene (one out-of-plane p orbital per carbon site with nearest-neighbor hopping energy of 2.8\,eV), $\phi$ is the self-consistent electric potential including both the impinging light and the Hartree interaction, and a phenomenological relaxation is assumed to bring the system to the relaxed state $\rho^0$ at a rate $\tau^{-1}$ with $\hbar\tau^{-1}=50$\,meV (i.e., the relaxation time is $\tau\approx13.2$\,fs). The density matrix $\rho=\sum_{jj'}\rho_{jj'}\ket{j}\bra{j'}$ is expressed in the basis set of one-electron eigenstates of $\HTB$, where $\rho_{jj'}$ are the sought-after time-dependent expansion coefficients. In particular, we have $\rho^0_{jj'}=\delta_{jj'}f_j$ for the relaxed state, where $f_j$ are Fermi-Dirac occupation numbers at the initial temperature $T=300\,$K. For ribbons, the states are treated as Bloch waves, arranged in bands as a function of their momentum along the direction of translational invariance, and the calculation is simplified by the orthogonality of different bands \cite{paper269}. The induced charge density at each carbon atom position $\Rb_l$ is then constructed as $\rho_l^{\rm ind}=-2e\sum_{jj'}\rho_{jj'}a_{jl}a^*_{j'l}$, where the factor of 2 originates in spin degeneracy, while the coefficients $a_{jl}$ represent the change of basis set between state $j$ and site $l$ representations. Finally, the time-dependent induced dipole and surface current are given by $\db(t)=\sum_l \Rb_l \rho^{\text{ind}}_l(t)$ and $\Jb(t)=\dot{\db}(t)$, respectively. For ribbons, we normalize these quantities per unit of ribbon length \cite{paper269}.

\subsection{MDF-CEM simulations}

In a complementary approach, we model electron dynamics in graphene within the MDF picture by adopting a non-perturbative semi-analytical model \cite{MCG16}, in which light-matter interaction is introduced through the electron quasi-momentum $\mbox{\boldmath${\pi}$} =\pb+(e/c){\bf A}$, where $\pb$ is the unperturbed electron momentum, ${\bf A}(t)=-c\int_{-\infty}^t\Eb(t')dt'$, and $\Eb$ is the classically-calculated in-plane electric field (see Sec.\ \ref{classical}). Electron dynamics is governed by the Dirac equation for massless fermions, which can be recast in the form of Bloch equations as \cite{I10,ASD14,MCG16}
\begin{widetext}
\begin{subequations}
\begin{align}
\dot{\Gamma}_\pb & =  - \left(\frac{1}{\tau} + 2i\omega_0\right)\Gamma_\pb - \frac{ie\,p_y}{p^2}\Eb(\Rb,t)\;n_\pb, \label{BEq1}\\
\dot{n}_\pb & = -\frac{1}{\tau} \left(n_\pb+1\right) + \frac{4e\,p_y}{p^2} \Eb(\Rb,t)\;{\rm Im}\left\{\Gamma_\pb\right\}, \label{BEq2}
\end{align}
\label{BEqs}
\end{subequations}
where $n_\pb(\Rb,t)$ and $\Gamma_\pb(\Rb,t)$ represent the population inversion and the interband coherence, respectively \cite{MCG16}. Here, the damping energy $\hbar\tau^{-1}=50$\,meV is the same as in the TB-RPA approach. These equations describe both inter- and intraband transitions. We solve Eqs.\ (\ref{BEqs}) nonperturbatively under the slowly-varying-envelope approximation \cite{MCG16} by expanding $\Gamma_\pb(\Rb,t) = \sum_{j=0}^N \left[ \Gamma_{\pb,j}^+(\Rb,t) e^{i(2j+1)\omega t} + \Gamma_{\pb,j}^-(\Rb,t) e^{-i(2j+1)\omega t} \right]$ and
$n_\pb(\Rb,t) = n_0(\Rb) + \sum_{j=1}^N {\rm Re}\left[ n_{\pb,j}^+(\Rb,t) e^{2ij\omega t} + n_{\pb,j}^-(\Rb,t) e^{-2ij\omega t} \right]$ in harmonic series up to $N=15$. The current is then parallel to the local electric field $\Eb(\Rb,t)\parallel\xx$, while its amplitude is calculated as an integral over momentum-resolved contributions,
\begin{align}
J(\Rb,t) = - \frac{ev_{\rm F}}{\pi^2\hbar^2} \int d^2\pb \left[\frac{p_x + e A}{\sqrt{(p_x + e A)^2+p_y^2}} (n_\pb+1) - \frac{2p_y}{p}\;{\rm Im}\left\{\Gamma_\pb\right\} \right]. \nonumber
\end{align}
Finally, the far-field power spectrum of the emitted light is proportional to $|\omega \langle\Jb(\Rb,\omega)\rangle|^2$, where $\Jb(\Rb,\omega)$ is the time-Fourier transform of $\Jb(\Rb,t)$, and $\langle\dots\rangle$ denotes the space average over the graphene structure under examination. 
\end{widetext}

\subsection{Classical electromagnetic simulations}
\label{classical}

The classical response of graphene nanostructures is simulated by numerically solving Maxwell's equations using the boundary-element method \cite{paper040} for ribbons and a finite-element method (COMSOL) for triangles. We describe graphene as a thin film (thickness $s=0.5\,$nm) and permittivity $\epsilon(\omega)=1+4\pi i\sigma(\omega)/\omega s$, where $\sigma(\omega)$ is the local-RPA conductivity \cite{paper235,GSC06,GSC09}. We thus obtain the linear optical extinction and the near-field distribution. Given the small lateral size of the ribbons and triangles compared with the light wavelength, we adopt a quasistatic eigenmode expansion \cite{paper228} and only retain one term corresponding to the dominant plasmon in each case. The incident light pulses are taken to have a large duration compared with the optical cycle, so we approximate them by a single carrier frequency (the pulse peak frequency) times a Gaussian envelope. We also use this approximation for the input near-field $\Eb$ of the MDF-CEM approach, with the carrier component classically calculated as explained above.

\section*{Acknowledgments}

We thank Jens Biegert and Fernando Sols for stimulating and enjoyable discussions and Renwen Yu for providing the plasmon wave function for triangles and the resonant near-field for ribbons. This work has been supported in part by the Spanish MINECO (MAT2014-59096-P and SEV2015-0522), AGAUR (2014 SGR 1400), Fundaci\'o Privada Cellex, and the European Commission (Graphene Flagship CNECT-ICT-604391 and FP7-ICT-2013-613024-GRASP).

%\bibliographystyle{apsrev}
%\bibliography{../../bibtex/refs}

\end{document}